# ANALYZING WEB SERVICES NETWORKS: A WS-NEXT APPLICATION

**Chantal Cherifi**, **Jean-François Santucci**
Corsica University, France
chantalbonner@gmail.com
santucci@univ-corse.fr

**ABSTRACT**

Web services represent a system with a huge number of units and many various and complex interactions. Complex networks as a tool for modelling and analyzing natural environments seem to be well adapted to such a complex system. To describe a set of Web services we propose three Web services network models based on the notions of dependency, interaction and similarity. Using the WS-NEXT extractor we instantiate the models with a collection of Web services descriptions. We take advantage of complex network properties to provide an analyzis of the Web services networks. Those networks and the knowledge of their toplogical properties can be exploited for the discovery and composition processes.

## 1 INTRODUCTION

A Web service is a set of related functionalities that can be published and discovered in a Web services registry and invoked for remote use. Those modular applications can be programmatically loosely coupled through the Web to form more complex ones. Two of the most popular problems in Web service technology addressed by both industry and academia are discovery and composition [1]. Discovery is the process of locating advertised Web services that can satisfy a service request. Composition arises when several Web services are needed to fulfill a request. The way those processes are achieved depends on how Web services are described. For syntactic Web services descriptions, discovery is performed on registries using keywords. Compositions are manually defined before any submitted request. Semantic descriptions allow automatic discovery and composition processes. Nevertheless finding the right Web services to fulfill a given request is not an easy task. Indeed, Web services are extremely volatile. Their number is continuously growing, and providers may change, relocate, or even remove them.

In this context, the Web services substitution play an important role within the composition process. Substitution consists in using a Web service instead of another. The only constraint is that the replacing one produces the same output and satisfies the same requirements as the replaced one. To perform Web services substitution, the Web services classification process aims at grouping Web services into categories usually called communities. Hence, works in Web services classification aim at grouping Web services according to some similarity criteria [2] [3] [4].

Classification is a step in structuring the Web services space to improve discovery and composition processes. Other criteria can be used to organize a set of Web services like their ability to be composed. In this case communities are formed with Web services that can interact in a composition.

On the one hand, Web services represent a system composed by a large number of highly interconnected dynamical units. On another hand, complex networks are a tool specifically dedicated to model such natural and complex systems. They allow studying their structure and dynamics [5]. Hence, a set of Web services can naturally be represented under the form of networks according to different criteria such as their similarity or their ability to be composed. Such kind of structures constitutes a convenient way to represent a set of Web services for visualization and analysis purposes. Moreover they can be stored and serve as a guide for Web services discovery and composition.

In this article, we introduce three models to structure a set of Web services. A dependency and an interaction model materialize Web services composition. A similarity model materializes similarity between Web services. We then provide a topological analysis of the networks structure using a well known benchmark.

The rest of the paper is organized as follows.

Background key elements are provided in section 2. Section 3 is dedicated to the literature review. Variables used to elaborate networks taxonomy are presented in section 4. In section 5 we introduce networks definitions. The networks taxonomies are presented in section 5. In section 6 we provide an analysis of the structure of some network samples. Finally, conclusions are provided in section 7.

## 2 BACKGROUND

In this section we give some background elements on Web services definition, Web services description languages, Web services discovery and composition and Web services classification.

### 2.1 Web Service Definition

Different kinds of information are linked to the notion of Web service. Some non functional properties (service provider, quality of service, service location) are present aside the Web service functionalities. In this work we focus on the functional aspect of Web services. Hence, we consider a Web service as an interface. A Web service interface is defined as a set of operations. An operation $i$ represents a specific functionality. It is characterized by one set of input parameters noted $I_i$, and one set of output parameters noted $O_i$. $I_i$ is the required information in order to invoke a Web service operation $i$. At the Web service level, the set of input parameters of a Web service $k$ is $I_k = \cup\, I_i$ and the set of output parameters $O_k = \cup\, O_i$. Fig. 1 represents a Web service numbered 1 with two operations.

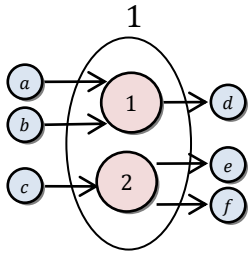

**Figure 1:** Schematic representation of a Web service 1, with two operations numbered 1 and 2. Operation 1: ($I_1 = \{a,b\}$, $O_1 = \{d\}$), Operation 2: ($I_2 = \{c\}$, $O_2 = \{e,f\}$), Web service 1: ($I_1 = \{a,b,c\}$, $O_1 = \{d,e,f\}$).

### 2.2 Description Languages

Production Web services are mostly expressed with Web Service Description Language (WSDL), a syntactic Web services description language [6]. This XML-based language has been proposed in the context of the W3C. More recently, the research community followed the current semantic Web trend by introducing semantics in Web services descriptions. Semantic Web services aim at augmenting Web services with rich formal descriptions of their capabilities. Several initiatives for semantic description languages exist and we can distinguish two main conceptual approaches. The first one aims at semantically annotating existing WSDL descriptions of Web Services. WSDL-Semantic (WSDL-S) [7] or Semantic Annotation for WSDL (SAWSDL) [8] are two semantic extensions of WSLD. The second approach aims at developing pure semantic Web services. The field includes substantial bodies of work, such as the efforts around Ontology Web Language for Services (OWL-S) [9]. OWL-S is an ontology of Web services specified by the W3C.

### 2.3 Composition

Web services composition addresses the situation when a request cannot be satisfied by any available atomic Web service. In this case, a composite Web service is synthesized to fulfill the request. A composite Web service is obtained by combining existing available atomic or even other composite Web services. The composition synthesis thus produces a specification of how to link the available Web services to realize the request.

### 2.4 Classification

Considering a set of Web services, the classification process aims at grouping them into categories. These categories are usually called communities. As in the literature classification is mainly performed according to the similarity between Web services, we will focus in work based on this definition. In this case there are two approaches to define communities i.e. top-down or bottom-up. In the former, abstract communities are designed a priori, and Web services are then defined in order to fit these categories [2] [3] [4]. In the later, communities are mined from an existing Web services collection [10] [11].

## 3 LITERATURE SURVEY

Despite the great potential they offer in terms of analysis tools, complex networks have not been widely used in the Web services area so far. Nevertheless, some authors already followed this recent trend to structure a set of Web services.

In [12] the authors define three composition network models according to the node types that can be parameters, operations or Web services. They use syntactic Web services to build networks considering either a full or a partial interaction between the nodes. Two types of syntactic matching i.e. equal and flexible are used to compute the links between networks nodes. Using complex network theory, they provide an analysis of the topological landscape of Web services networks formed by a real-world data set.

In [13] and [14] the authors provide an interaction network model with Web services as nodes. They deal with semantic Web services considering a partial interaction mode. Equivalence and subsumption ontological concepts relationships are considered to compute the links between networks nodes. In [13], complex network theory is also used to rank the Web services according to their connectivity. Experiments are performed on an automatically generated and simulated Web services network. A composition algorithm is applied to the networks while being dynamically guided by the ranking. In [14] the network is built from a set of artificial Web services descriptions. To synthesize a composition plan the network is explored with a backward chaining discovery.

In [15] the authors propose a dependency network model with parameters as nodes. The model is based on semantically described Web services. The network is used to derive composite Web services with a breadth first search algorithm.

In [16] an interaction Web services network is proposed. The goal of this work is to classify Web services. The authors provide a graph based method for composition oriented Web services classification using a b-coloring approach.

From all these works we can observe that there are various ways to represent a Web services set as a network. We can identify some variables to distinguish the proposed models. The Web services description, the network nodes, the relationship between nodes, the amount of information considered to establish a relation between two nodes and the matching are among the variables that allows building different types of networks. To evaluate all the possible models based on these variables we derived a tool, WS-NEXT, that allows building associated networks from a given set of Web services [17].

## 4 NETWORK VARIABLES

In this section we give an accurate meaning of the previously identified Web services network variables that can be used to modulate a Web services network.

### 4.1 Description
The description variable represents the Web service description type. Those two types are syntactic and semantic descriptions. Corresponding variables values are respectively noted *syntactic* and *semantic*. In a syntactic description, each parameter has a name and an XML type. In a semantic description the name and the type are also generally specified and an additional ontological concept is associated to the parameter. Ontological concepts are domain specific and consensual terms. They give parameters a contextual and precise meaning.

### 4.2 Granularity
The granularity determines the nodes entities i.e. the nature of the nodes in a network. From coarser to finer, we consider Web services, operations or parameters as node entities. We note the corresponding variables values as *service*, *operation* and *parameter*.

### 4.3 Model
The model expresses the nature of the links i.e. the type of relationship between nodes. This relationship depends on the granularity.

Considering parameters as nodes, if one is an input parameter of a Web service (or of an operation) and if the other is an output parameter of the same Web service (or operation), there is a dependency relationship between them. Indeed, the production of the second parameter depends on the provision of the first one through the invocation of the Web service (or of the operation). This model is noted *dependency* and is illustrated by Fig. 2. One Web service numbered 2 is considered with one operation numbered 3 and its parameters as follows: 3 ($I_3 = \{f\}$, $O_3 = \{g, h\}$). Parameters $h$ and $g$ depends on the provision of parameter $f$, hence there is a dependency relationship between them $f$ and $h$ and $g$.

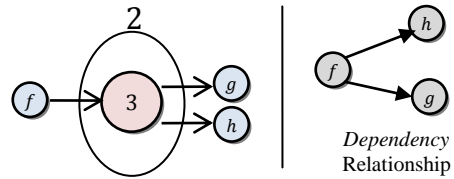

*Dependency* Relationship

**Figure 2:** Example of the *Dependency* Model. Right side: one Web service with one operation and its parameters. Left side: corresponding dependency relationships between parameters.

Considering Web services or operations as nodes, a relationship between two nodes corresponds to the information flow between them. In other words the first one is able to provide the information needed by the second one in order to invoke it. This model is called *interaction*. It is illustrated by Fig. 3. Two Web services 1 and 2 are considered. Web service 1 has two operations as follows: 1 ($I_1 = \{a, b\}$, $O_1 = \{d\}$) and 2 ($I_2 = \{c\}$, $O_2 = \{e, f\}$). Web service 2 has one operation as follows: 3 ($I_3 = \{f\}$, $O_3 = \{g, h\}$). Web service 1 can provide the information in order to invoke Web service 2 , hence there exist an interaction relationship between them. Operation 2 can provide the information in order to invoke operation 3 , hence there exist an interaction relationship between them.

Dependency and interaction models are different ways to materialize Web services composition.

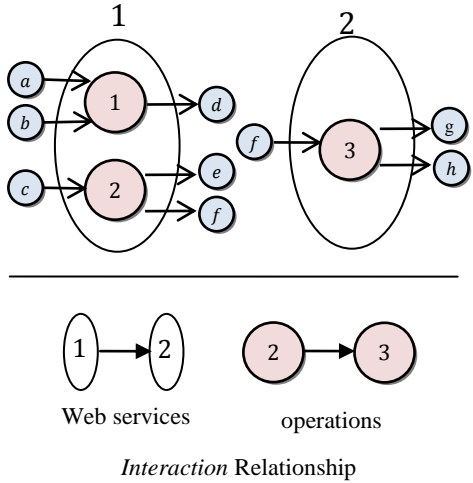

*Interaction* Relationship

**Figure 3:** Example of the *Interaction* Model. Top: two Web services with their respective operations and parameters. Bottom: corresponding interaction relationships between Web services and between operations.

Considering operations as nodes, a relationship between two operations corresponds to a certain type of similarity between them. The similarity relation can be either symmetrical or asymmetrical. In the first case, the two operations are said to be similar to each other. In the second case, the second operation is said to be similar to the first one according to some criteria. This model is noted *similarity* and is illustrated by Fig. 4. Three operations with their respective parameters are considered as follows: 1 ($I_1 = \{a,b\}$, $O_1 = \{d\}$), 4 ($I_4 = \{a\}$, $O_4 = \{b,d\}$), 5 ($I_5 = \{a,b\}$, $O_5 = \{d,e\}$). The similarity relation between operations 1 and 4 is symmetrical.

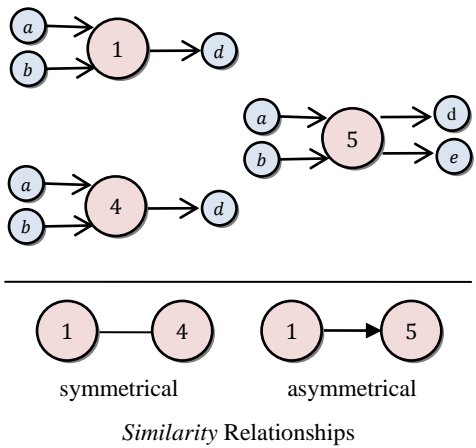

*Similarity* Relationships

**Figure 4:** Example of the *Similarity* Model. Top: three operations with their respective parameters. Bottom: corresponding similarity relationships between operations.

Indeed the two operations are symmetrically similar because they have the same output parameters. An asymmetrical similarity relationship exists between operation 1 and operation 5. Indeed, operation 5 has more output parameters than operation 1.

### 4.4 Mode

The mode represents the amount of information used to link two nodes in a network. This variable is related to the interaction model. Two cases must be considered. Either all the information is provided or only part of this information exists. If a Web service or an operation can provide all the parameters values needed to invoke another one, we will denote this case as *full* interaction mode. Fig. 3 is an example of the full interaction mode. A full interaction exists between Web service 1 and Web service 2. Indeed, Web service 2 needs only parameter f to be invoked and Web service 1 can provide this information. A full interaction also exists between operation 1 and operation 3. If a Web service or an operation cannot provide all the input parameters required by a second one, this mode is denoted by *partial*. Such a case is illustrated by Fig. 4. Two Web services 2 and 3 are considered. Web service 2 has one operation 3 ($I_3 = \{f\}$, $O_3 = \{g,h\}$). Web service 3 has one operation 6 ($I_6 = \{g,h,i\}$, $O_6 = \{j,k\}$). There is a partial interaction between Web service 2 and Web service 3. Indeed, Web service 2 can provide only part of the information needed by Web service 3 which is parameters $g$ and $h$.

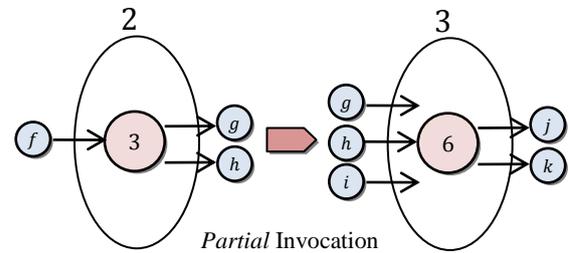

*Partial* Invocation

**Figure 5:** Example of *Partial* Interaction Mode.

### 4.5 Matching

The matching variable is related to the similarity measures between parameters. It is computed differently for syntactic and semantic descriptions. For syntactic descriptions, matching consists of comparing two Web services parameters names using similarity metrics. We distinguish two cases. The first case considers two parameters as similar if their names are exactly the same string. It is called *equal*. The second case considers two parameters as similar if their name presents a certain level of similarity. It is called *approximate*. Different similarity metrics can be used. Classical similarity metrics have been considered in WS-NEXT (Levenshtein, Jaro and Jaro-Winkler). These metrics are denoted as *Levenshtein*, *Jaro* and

*Winkler*. We also developed a smoothed metric based on Levenshtein distance between filtered strings denoted *Smoothed*.

For semantic descriptions, matching consists in comparing ontological concepts associated to the parameters. This is done by the classical operators (exact, plugin and subsume) that have been developed in previous work in the matchmaking area [18]. Exact corresponds to a perfect matching, i.e. both concepts belong to the same ontology and are exactly identical. When the concept associated to the first parameter is strictly more specific than the other one, plugin is used. Subsume is used when the first concept is strictly more general than the second one. We add a fourth matching operator called fitin which consider the case where there is simultaneously plugin and exact similarities between two nodes. This operator leads to a more flexible semantic interaction representation. The matching variables values are denoted by *exact*, *plugin*, *subsume* and *fitin*.

## 5 NETWORKS DEFINITION

Dependency, interaction and similarity networks can be used to represent a set of Web services. In dependency networks nodes are parameters while interaction networks can use either operations or Web services as nodes. In similarity networks nodes are operations. In the following we define the three corresponding network models.

### 5.1 Dependency Network

A dependency network is defined as a directed graph whose nodes correspond to depending parameters and links indicate the head parameter depends on the tail parameter (as illustrated by Fig. 2, g depends on f) [19]. In the context of dependency networks, each Web service $w$ is formally defined as a triplet $(I_w, O_w, K_w)$. $K_w$ denotes the set of dependencies defined by $w$. Each operation $i$ is formally defined as a triplet $(I_i, O_i, K_i)$. $K_i$ denotes the set of dependencies defined by $i$. We consider each output parameter depends on each input parameter. To build such a network, we first associate one node to each parameter present in the whole collection. Then, links are drawn by considering each Web service (or operation) separately. A link is added between each one of its input parameters and each one of its output parameters. Additionally, one parameter may be used by several Web services or operations, either as an input or an output. Consequently, we have to decide if two parameters are similar. This is done trough the matching functions described in section 4. In the case of syntactic dependency network, *equal* matching is applied. For a semantic description *exact* matching is applicable.

### 5.2 Interaction Network

We define an interaction network as a directed graph whose nodes correspond to interacting Web services and links indicate the possibility for the tail Web service to act on the head Web service [20]. To represent a collection of Web services descriptions as an interaction network of Web services, we first define a node to represent each Web service in the collection. Then, a link is drawn from a Web service 1 to a Web service 2 if for each input parameter in $I_2$, a similar output parameter exists in $O_2$. In other words, the link exists if and only if Web service 1 can provide the information requested to invoke Web service 2. In the interaction network, a link between two Web services therefore represents the possibility to compose them. Similarly, we can define an interaction network at the operation level. The matching functions described in section 4 are used to determine the similarity between two parameters.

### 5.3 Similarity Network

We define a similarity network [21] as a graph whose nodes correspond to possibly similar Web services operations. To represent a collection of Web services as a similarity network of operations, we first associate a node to each operation in the collection. Then, a link is added between two nodes if the corresponding operations are similar. The similarity relation between two sets of parameters can be approached in several ways. To that end, we defined four similarity functions. They are respectively named Full Similarity (*FullSim*), Partial Similarity (*PartialSim*), Excess Similarity (*ExcessSim*) and Relation Similarity (*RelationSim*). These functions are defined in terms of set relations between the input and output parameters sets of the compared operations. Let $I_i$ and $O_i$ be the sets of input and output parameters for operation $i$ respectively. Suppose we want to compare operation 1 and operation 2. FullSim states both operations are fully similar if they provide exactly the same outputs ($O_1 = O_2$) and if they have overlapping inputs ($I_1 \cap I_2 \neq \emptyset$). PartialSim states 2 is partially similar to 1 if some outputs of 1 are missing in 2 ($O_1 \supset O_2$) and if they have overlapping inputs ($I_1 \cap I_2 \neq \emptyset$). ExcessSim states 2 is similar to 1 with excess if 2 provides all the outputs of 1 plus additional ones ($O_1 \subset O_2$) and if 2 has at most the inputs of 1 ($I_1 \supseteq I_2$). The RelationSim function states both operations have a relational similarity if they have exactly the same outputs ($O_1 = O_2$) and if they do not share any common input ($I_1 \cap I_2 = \emptyset$). To determine the relations between two sets of parameters, one needs to be able to compare the parameters themselves. Hence, the similarity functions are based on the equal matching described in section 4.

## 6  WEB SERVICES NETWORKS

In order to build Web services networks from a set of Web services descriptions, we used WS-NEXT (Web Services Network Extractor). WS-NEXT allows building networks from a collection of Web services descriptions files, according to the network models and the variables previously defined. Networks that can be extracted by WS-NEXT are figured by a tree starting from the root, going through each variable and ending by an underlined leaf.

### 6.1  Dependency Taxonomy

Fig. 6 shows the dependency networks taxonomy. Two dependency networks can be extracted with WS-NEXT, one syntactic with equal matching and one semantic with exact matching.

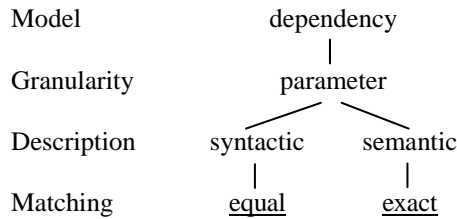

**Figure 6:** Dependency Networks. Left side: network variables. Right side: networks.

### 6.2  Interaction Taxonomy

The interaction networks taxonomy is depicted by Fig. 7 and Fig. 8. Eighteen full interaction networks and eighteen partial interaction networks can be extracted with WS-NEXT.

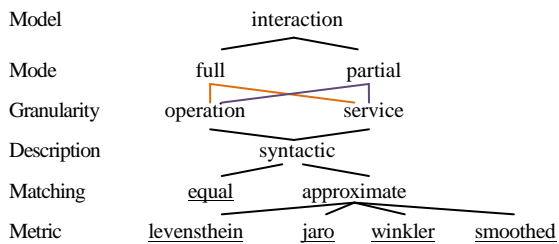

**Figure 7:** Syntactic Interaction Networks. Left side: network variables. Right side: networks.

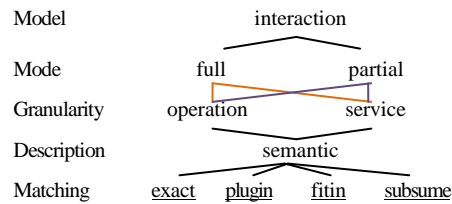

**Figure 8:** Semantic Interaction Networks. Left side: network variables. Right side: networks.

### 6.3  Similarity Taxonomy

The taxonomy of similarity networks is represented on Fig. 9. We can extract eight similarity networks with WS-NEXT.

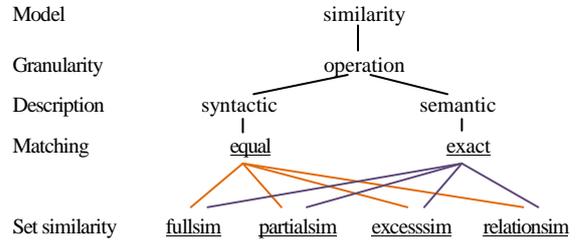

**Figure 9:** Similarity Networks Taxonomy. Left side: network variables. Right side: networks.

## 7  NETWORKS EXTRACTION AND ANALYSIS

From a collection of Web services descriptions, we extracted a set of ten networks as follows. Two dependency networks: 1 syntactic (equal) and 1 semantic (exact); four interaction networks: 1 syntactic (equal) and 3 semantic (1 exact, 1 plugin, 1 subsume); four syntactic similarity networks: 1 FullSim, 1 PartialSim, 1 ExcessSim, 1 RelationSim. We then investigated the structural properties of the networks.

In these experiments, we follow two main objectives. First, we want to study the influence of semantics on the composition process by comparing the structure of the syntactic and semantic composition networks. Second, we want to investigate the relation between the network structure and the application for which it is devised. Usually, networked systems exhibit a component organization. A component is a maximal connected sub-graph disconnected from the rest of the network. Either a network exhibit a giant component with small other ones, or all the components have an equivalent size. By tracking the component organization, the components size and links number, we can interpret the analysis results in terms of Web services composition and substitution.

The networks have been extracted from the SAWSDL-TC1 [22] Web services descriptions collection. Indeed, in this work, we want to simulate real-world conditions and to compare syntactic and semantic composition Web services networks. Hence, one need to have a collection of a large number of real-world Web services described both syntactically and semantically. SAWSDL-TC1 provides 894 Web services descriptions written in SAWSDL. Each description contains only one operation. The collection contains 2136 parameter instances. Parameters are syntactically described by their name and semantically described by their ontological concept. The descriptions are distributed over 7 thematic domains (education, medical care, food, travel, communication, economy and weapon). The collection originates in the OWLS-TC2.2 collection, which contains a part of real-world Web services descriptions retrieved from public IBM UDDI registries, and semi-automatically transformed from WSDL to OWL-S.

### 7.1 Structure of Dependency Networks

We extracted the syntactic and the semantic dependency networks with WS-NEXT, according to the dependency network definition. As matching functions gather similar parameters, there is a significant difference between the number of instances in the collection and the number of parameter nodes in the networks. The 2136 parameters instances of the collection are represented by 385 nodes in the syntactic network and by 357 nodes in the semantic one. As we used different matching functions to build the syntactic and semantic networks, the sets of similar parameters are not the same in the syntactic network and in the semantic network. The number of nodes is smaller in the semantic network. This indicates semantic matching allows associating more parameter instances. This result highlights the presence of false negatives in the syntactic network. False negatives are instances associated to different nodes in the dependency network. They are actually conveying the same information and should be represented by the same node. These false negatives are usually syntactically different because they have different names. But they have the same meaning, hence they are associated to the same ontological concept. For example parameter instances `_AUTHOR`, `_AUTHOR1` and `_AUTHOR2` are represented by three distinct nodes in the syntactic network. In the semantic network, they are associated to a unique node as they all are associated to the same `#author` concept. The semantic matching also allows eliminating some false positives. False positives correspond to instances represented by the same node whereas they do not represent the same information. For example, many instances are simply called `PARAMETER` but are associated to very different concepts. The syntactic matching will improperly associate them to a common node, whereas the semantic matching will not.

Both networks exhibit the same structure: a giant component along with several small components and isolated nodes. Nevertheless, the distribution between these three types of entities is slightly different. The proportion of isolated nodes is 4.67% in the syntactic network and 4.2% in the semantic network. While this value is smaller for the semantic network, the number of isolated nodes remains quite small in both networks. Isolated nodes are parameters belonging to Web services having only input parameters or to Web services having only output parameters, and they are exclusively either input or output. The giant component in the syntactic network contains 73% of the remaining nodes and 86% of the remaining links while in the semantic networks it contains 78% of the nodes and 88% of the links. The syntactic network exhibits 17 other smaller components with a size ranging from 2 to 30 nodes. This is to compare to 15 small components for the semantic network (2 to 14 nodes). Fig. 10 shows the trimmed semantic network (isolated nodes have been discarded). The giant component stands in the middle surrounded by the small components.

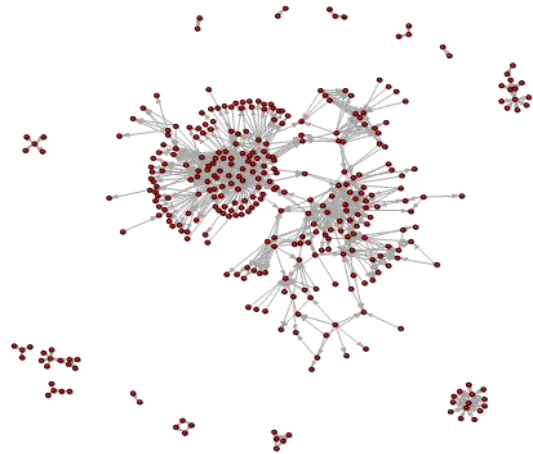

**Figure 10:** Trimmed exact semantic dependency network. The giant component is located in the middle surrounded by small components.

The semantic network presents less isolated nodes and less small components than the syntactic one. These properties are more effective in terms of composition ability. Recall that if many distinct components exist this reflects that the collection is made of several non-interacting groups of parameters. Furthermore the semantic network has a larger giant component than the syntactic one both in terms of nodes and links. It shows that the numbers of dependencies in which several operations are implied is higher. These results demonstrate that a larger proportion of Web services can interact if one uses the semantic network.

### 7.2 Structure of Interaction Networks

According to the interaction network definition, we extracted 4 networks with WS-NEXT from SAWSDL-TC1 collection, one syntactic and three semantic. The syntactic network is the full equal network. In some previous work [23] we performed a comparative study on the metrics performance by studying the topological properties of syntactic approximate networks. It appears that the use of the approximate metrics to build interaction networks is not very satisfying. For this reason, we concentrate on the equal network. The semantics networks are the full exact, the full plugin network and the full subsume network. In this study, we discarded the fitin network to keep and compare only strict subsumption relationships. We restrict our investigations to the full mode. Indeed, we want to put ahead eventual differences between syntactic and semantic network structures not to compare intra-model variations.

The number of nodes and links is globally higher in the syntactic network than in the semantic networks. The syntactic network contains 395 nodes and 3666 links. The exact network contains 341 nodes and 3426 links. The plugin network contains 369 nodes and 2446 links. The subsume network contains 329 nodes and 3864 links. This result is the consequence of the presence of some false positives in the syntactic network.

The same structure is shared by all the networks. We remark the presence of isolated nodes, a giant component and small components much smaller than the giant one. The four networks contain many isolated nodes. They represent 44% of the total nodes in the syntactic network. This proportion is approximately 49% in the semantic networks. There is less isolated nodes in the syntactic network because some nodes are inappropriately linked to others and cannot participate in a composition. In an interaction network, isolated nodes represent Web services that do not interact with others. None of their output parameter can serve as input and none of their input parameter is provided by other Web services. Hence, they only can be invoked as atomic Web services. In the syntactic network, the giant component contains 90% of the remaining nodes and 99% of the remaining links. The giant component of the exact network contains 85% of the nodes and 98% of the links in the trimmed network. The plugin and the subsume semantic networks present very similar proportions. Once again these results highlight the presence of false positives in the syntactic network. The syntactic network exhibit 5 small components ranging from 2 to 22 nodes. The exact network has 7 small components whose sizes range from 2 to 28. Plugin and subsume networks exhibit 5 small components respectively ranging from 3 to 10 and from 5 to 90. Fig. 11 shows the trimmed exact semantic network separated in 8 components. The small components are less numerous and smaller in the syntactic network because of the presence of false positives that have been integrated in the giant component. In the plugin and in the subsume networks, the constraints on the interactions are relaxed comparing to the exact network. Hence, nodes are gathered within fewer components.

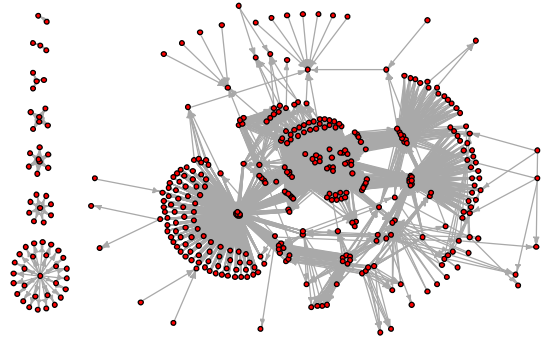

**Figure 11:** Trimmed exact semantic interaction network. The giant component is located on the right side. The small components stand in the right side.

The exact semantic network presents more isolated nodes, more small components and a smaller giant component than the syntactic one. These properties seem to be less effective in terms of composition ability. Nevertheless, the interconnection structure is more accurate in a semantic network. It should consequently results in a more efficient composition discovery process. One may consider the plugin and the subsume networks as additional solutions for this task. In this case, the resulting semantic search space becomes larger than the syntactic one.

### 7.3 Structure of Similarity Networks

According to the definitions, four syntactic similarity networks have been extracted with WS-NEXT from SAWSDL-TC1 collection. We choose to study only one description type to concentrate on similarity functions comparison. The four networks contain 785 nodes, corresponding to the 785 operations of the collection. Table 1 summarizes the values of the networks properties.

**Table 1:** Properties of the full, partial, excess and relational similarity networks.

| Property | Full Sim | Partial Sim | Excess Sim | Relation Sim |
|---|---|---|---|---|
| **Isolated nodes** | 604 | 447 | 486 | 227 |
| **Nodes in trimmed network** | 181 | 338 | 299 | 548 |
| **Components** | 38 | 61 | 67 | 123 |
| **Links** | 310 | 412 | 307 | 2254 |

Except for the first row, all the others properties are computed on the trimmed networks, i.e. without any isolated nodes. For all similarity networks under study no giant component is emerging, but numerous small ones, along isolated nodes. This reflects the decomposition of the collection into a reasonable number of communities. This is a good thing, because having only isolated nodes or a giant component would lead to useless communities. Indeed, in the former case, each community would contain only one operation, and in the latter all operations would be considered as similar to the all others. Both cases would have been surprising considering we processed a real-world collection. To illustrate the structure of the similarity network, a typical component from the partial network is presented in Fig. 12.

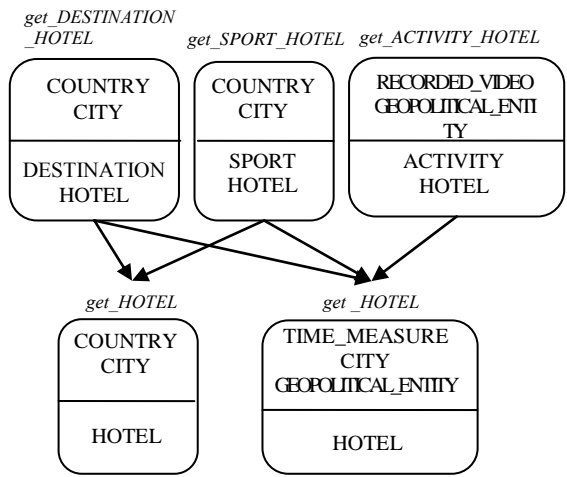

**Figure 12:** A component of the partial similarity network with 5 nodes.

Operations `get_DESTINATION_HOTEL`, `get_SPORT_HOTEL`, `get_ACTIVITY_HOTEL` are linked with `get_HOTEL`. Indeed `get_HOTEL` operation provides only the `HOTEL` output parameter while the three others provide the `HOTEL` output parameter and an additional specific one. A `get_HOTEL` operation can partially satisfy a destination/hotel request, an activity/hotel request or a sport/hotel request.

From full similarity to relation similarity according to table 1 order, the number of isolated nodes globally decreases while the number of links and components increases. Indeed, as constraints on outputs become less strict, more links are created leading to new components or to the increase of the population of the existing ones. The number of nodes, the number of links and the number of components are the highest in the relation similarity network. In this collection, a lot of operations produce identical outputs with completely different inputs.

Let's inspect the number of components containing 90% of nodes all together. We need 17 components in the full similarity network, 30 in the partial similarity network, 40 in the excess similarity network and 32 in the relation similarity network. Those results show that at least half of the components contain very few nodes while the other half contains at least 90% of nodes. These small components are not very interesting; they do not offer many opportunities in terms of substitution.

## 8 DISCUSSION

From the comparison between syntactic and semantic networks, for both dependency and interaction models, it appears that the semantics in the Web services descriptions leads to more accurate interconnection structures. Indeed, we demonstrated that the inappropriate dependencies and interactions relationships that appear in the syntactic networks are discarded in the semantic networks due to the use of ontologies and semantic matching. One can expect though, a more efficient composition process using the semantic description. A large body of work exists in the domain of semantic descriptions and automatic Web services composition. Nevertheless, production Web services still widely rely on syntactic descriptions. To take advantage of the great potential of a semantic Web services pool, one should be able to annotate legacy Web services descriptions. Manual annotation is a complex and costly task hence there is a need to appropriate annotation tools. Few researchers have proposed solutions for this task [24] [25]. At this point there is no satisfying solution that can perform an efficient fully automatic annotation. Bridging the gap between a syntactic and a semantic notation is a difficult problem. We suggest devising semi automatic annotation tools as a first step towards this goal.

The giant component in the composition networks structure reflects the presence of a huge number of interconnected Web services. In these networks, the presence of a giant component is of great importance. It represents the largest fraction of the network within which compositions are possible. It is a guaranty for a composition process to be successful. In the similarity networks, no giant component emerges. They are rather divided into numerous small components. This structure reflects the decomposition of the networks into many Web services communities and, as a consequence, of substitutable operations. A composition process could take advantage of these two complementary structures. We can combine the two structures to obtain a two-level architecture. We suggest an upper level containing an interaction network. Each node of this network could be an abstract operation gathering similar concrete operations. Hence, the composition search space would be reduced. The lower level is then represented by the similar networks. The abstract operations of a composition could be instantiated by concrete operations of the lower level with the possibility of substitution.

## 9 CONCLUSION

In this paper, we proposed three network models to structure a set of Web services. The models aim at organizing the Web services according two different directions. The first one takes into account the composition relationship between Web services. The second one is based on their similarity relationships.

We provide a topological analysis of the networks. This analysis shows that the structure of the semantic description for composition networks is more accurate. Networks formed from the composition models exhibit a giant component in which a large number of Web services are interacting. Networks formed from the similarity model are composed by many small components which gather a pool of similar operations.

Our future work will focus on algorithms based on the composition and similarity networks for composition discovery and substitution purpose.